\documentclass[a4paper, twocolumn, 11pt]{quantumarticle}
\pdfoutput=1
\usepackage[utf8]{inputenc}
\usepackage[T1]{fontenc}
\usepackage{amsmath, amssymb, amsthm}
\usepackage{graphicx}
\usepackage{hyperref}
\usepackage{braket}
\usepackage{qcircuit}
\usepackage{physics}
\usepackage{booktabs}
\usepackage{multirow}
\usepackage[numbers,sort&compress]{natbib}
\usepackage{needspace}
\usepackage{listings}
\usepackage{xcolor}
\lstdefinestyle{qsg}{
  language=Python,
  basicstyle=\ttfamily\small,
  keywordstyle=\color{blue},
  commentstyle=\color{gray},
  stringstyle=\color{red},
  showstringspaces=false,
  breaklines=true,
  columns=fullflexible,
  frame=single,
  captionpos=b
}
\title{Quantum Skip Gates: Coherently Conditioned Subroutines in Iterative Quantum Algorithms}
\author{Kym Derriman}
\affiliation{Department of Physics and Astronomy, Rutgers University, Piscataway, NJ, USA}
\email{kym.derriman@rutgers.edu}
\date{January 23, 2026}
\begin{document}
\maketitle
\begin{abstract}
The Quantum Skip Gate (QSG) is a unitary circuit primitive that coherently superposes execution and omission of an expensive quantum subroutine based on the outcome of a cheaper preceding subroutine, without mid-circuit measurement or loss of coherence. By using a control qubit and an internal flag, QSG enables conditional quantum logic in a fully unitary setting. Demonstrated experimentally in a Grover-style search on IBM quantum hardware ($n=4$, $k=3$), the QSG reduces costly subroutine calls by $9$--$25\%$, achieving $31$--$61\%$ higher success-per-oracle efficiency relative to a fixed-order baseline. Noise-model simulations confirm and strengthen these gains (up to $45\%$) when using an optimized ``swap-out'' design. These results demonstrate that coherently conditioned subroutines provide practical resource management, significantly reducing runtime costs and noise accumulation in near-term quantum algorithms.
\end{abstract}

\section{Quantum Skip Gates}
Quantum algorithms often contain expensive subroutines whose execution can be avoided if an earlier, cheaper subroutine has already succeeded. Classically, this would require measurement and feed-forward control, breaking coherence. The Quantum Skip Gate (QSG) is a unitary primitive that enables coherent superposition over whether an expensive subroutine $U_B$ is applied or skipped, based on the result of a preceding inexpensive subroutine $U_A$.

A control qubit $C$ is prepared in superposition. After $U_A$ sets a flag $f_A$, the conjunction $C \land f_A$ is coherently computed into an ancilla $a$. This ancilla then controls whether $U_B$ acts on the data register (via multi-controlled gates or controlled-SWAP redirection). The ancilla is uncomputed, restoring full coherence. In the branch where $C=1$ and $f_A=1$, $U_B$ is coherently replaced by the identity.

The QSG thus implements a coherent superposition of ``run'' and ``skip'' branches entirely within a unitary circuit. The swap-out variant (detailed later) avoids depth explosion when making deep subroutines conditional. This construction is broadly applicable to iterative quantum algorithms where early termination or conditional execution can save resources.

\begin{figure}[tb]
    \centering
    \includegraphics[width=\linewidth]{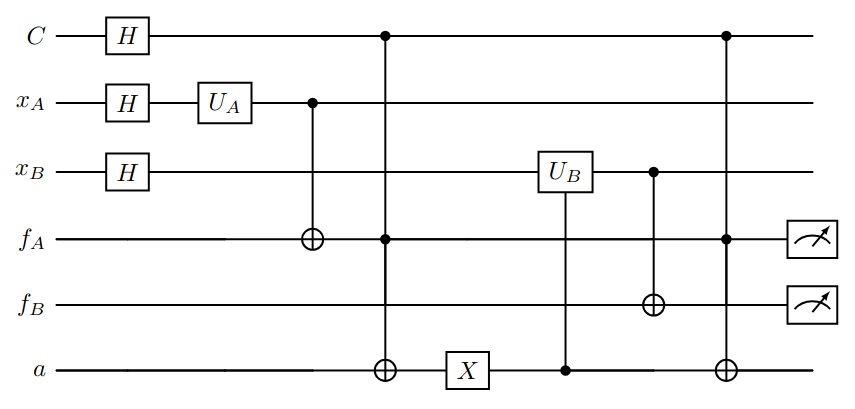}
    \caption{Generalized structure of the quantum skip gate (QSG). The control qubit $C$ and data registers $x_A$, $x_B$ are initialized in superposition. A subroutine $U_A$ marks $x_A$, and a controlled Toffoli-style gate encodes the skip condition $C \land f_A$ into an ancilla $a$. The subroutine $U_B$ is applied to $x_B$ only if $a = 0$ and skipped when $a = 1$. The ancilla is uncomputed to restore coherence. The depth-optimized swap-out realization is omitted.}
    \label{fig:highlevel-qsg}
\end{figure}

\section{Formal Description of Quantum Skip Gate (QSG)}
\subsection{Register Structure}
We consider a composite quantum system defined over the Hilbert space
\begin{equation}
\mathcal{H} = \mathcal{H}_C \otimes \mathcal{H}_{x_A} \otimes \mathcal{H}_{x_B} \otimes \mathcal{H}_{f_A} \otimes \mathcal{H}_{f_B},
\end{equation}
where $\mathcal{H}_C$ is the two-dimensional space for a control qubit $C$, $\mathcal{H}_{x_A}$ and $\mathcal{H}_{x_B}$ are $2^n$-dimensional spaces corresponding to two $n$-qubit data registers, and $\mathcal{H}_{f_A}$ and $\mathcal{H}_{f_B}$ are ancillary spaces used to store internal flags. These flags may encode intermediate results used to condition subsequent operations. In some applications, like a Grover-style search, they may indicate whether a prior operation has already succeeded. If a swap-based implementation of skip logic is used, we also include an auxiliary register $d_B$ with Hilbert space $\mathcal{H}_{d_B} \cong \mathcal{H}_{x_B}$, which is initialized to $\ket{0}^{\otimes n}$ at the start of each iteration and remains disentangled at the end. We do not analyze the swap-based construction in this section, as it is not essential to understanding the logic of the Quantum Skip Gate; it is noted here solely for completeness.

\subsection{Initial State Preparation}
At the start of the circuit, the control qubit $C$ is prepared in the state $\ket{+}_C = (\ket{0} + \ket{1})/\sqrt{2}$ to enable coherent branching. Each data register $x_A$ and $x_B$ is initialized to a uniform superposition over all $2^n$ basis states:
\begin{align}
\ket{+}^{\otimes n}_{x_A} &= \frac{1}{\sqrt{2^n}} \sum_{x \in \{0,1\}^n} \ket{x}_{x_A}, \nonumber \\
\ket{+}^{\otimes n}_{x_B} &= \frac{1}{\sqrt{2^n}} \sum_{y \in \{0,1\}^n} \ket{y}_{x_B}.
\end{align}
The flag qubits $f_A$ and $f_B$ are each initialized to $\ket{0}$, and any auxiliary register $d_B$ introduced for swap-out constructions is likewise initialized to $\ket{0}^{\otimes n}$. The full initial state of the system is thus
\begin{equation}
\begin{aligned}
\ket{\psi_0}
&= \ket{+}_C
   \otimes \ket{+}^{\otimes n}_{x_A}
   \otimes \ket{+}^{\otimes n}_{x_B} \\
&\quad \otimes \ket{0}_{f_A}
   \otimes \ket{0}_{f_B}
   \otimes \ket{0}^{\otimes n}_{d_B},
\end{aligned}
\end{equation}
where the final factor may be omitted if no swap-out mechanism is used. This initialization places the system into a superposition over control and data configurations, ready for coherently conditioned operations to follow.

\subsection{Application of Inexpensive Subroutine and Flag Encoding}
The first operation applied is a unitary subroutine $U_A$ acting on the register pair $(x_A, f_A)$. Its role is to coherently detect a condition on the contents of $x_A$ and encode the result into the flag qubit $f_A$. We model $U_A$ as a controlled bit-flip conditioned on a predicate $\Pi_A$, where $\Pi_A$ is a projector acting on $\mathcal{H}_{x_A}$ that defines the subroutine’s success condition. Explicitly, we write
\begin{equation}
\begin{aligned}
U_A
&= I_C \otimes \Big[ (I - \Pi_A) \otimes I_{f_A} \\
&\qquad\quad + \Pi_A \otimes X_{f_A} \Big]
\otimes I_{x_B f_B d_B},
\end{aligned}
\end{equation}
where $X_{f_A}$ is the Pauli-$X$ acting on the flag qubit, and the identity acts on all other registers. This operation flips $f_A$ from $\ket{0}$ to $\ket{1}$ if and only if the state of $x_A$ lies in the support of $\Pi_A$. The operator $U_A$ is fully unitary and does not disturb the quantum coherence of the system.

\subsection{Conditional Skip Logic}
To enable conditional omission of the second subroutine, we introduce an ancilla qubit $a$ initialized to $\ket{0}$ and use it to compute the logical conjunction of the control qubit $C$ and the flag qubit $f_A$. The skip condition is defined to hold precisely when $C = 1$ and $f_A = 1$, in which case the operation $U_B$ will be bypassed. Toffoli-style control logic is used to compute this condition coherently. The corresponding transformation maps the computational basis state as
\begin{equation}
\ket{C} \ket{f_A} \ket{0}_a \mapsto \ket{C} \ket{f_A} \ket{C \wedge f_A}_a,
\end{equation}
entangling the ancilla with the control–flag subsystem in a fully coherent manner, without measurement. In our implementation, this logic is realized using the RCCX construction, a relative-phase Toffoli gate that reduces circuit depth while preserving the intended classical behavior. The full operation is represented by a unitary $V_{\text{AND}}$ acting on the tuple $(C, f_A, a)$ and extended trivially to the rest of the system:
\begin{equation}
V_{\text{AND}} = \text{Toffoli}_{C, f_A \rightarrow a} \otimes I_{x_A x_B f_B d_B}.
\end{equation}
After this operation, the ancilla qubit $a$ stores the skip condition $C \wedge f_A$ and will serve as the control for the application (or omission) of the subroutine $U_B$ in the next stage of the circuit.

\subsection{Controlled Application of $U_B$}
Once the ancilla qubit $a$ encodes the conjunction $C \wedge f_A$, we apply an $X$ gate to invert its value, so that it now represents the negated condition $\lnot(C \wedge f_A)$ used to control the application of $U_B$. The second subroutine $U_B$ is then applied to register $x_B$ under the control of $a$. Formally,
\begin{equation}
\begin{aligned}
U_B^{\text{cond}}
&= \left( X_a \right) \\
&\quad \cdot \left(
      \ket{0}\bra{0}_a \otimes U_B
      + \ket{1}\bra{1}_a \otimes I
      \right) \\
&\quad \cdot \left( X_a \right),
\end{aligned}
\end{equation}
where $U_B$ acts on $x_B$ and the identity $I$ acts elsewhere. This ensures that $U_B$ is applied when $a = 0$, and skipped when $a = 1$, coherently across the superposition branches.

In practice, implementing $U_B$ in this form requires placing every gate inside $U_B$ under control by $a$, resulting in a multi-controlled subroutine with depth and resource overhead proportional to the size of $U_B$. For deep oracles or hardware with limited native multi-qubit gates, this approach becomes prohibitively expensive. To address this, we introduce a depth-optimized alternative using a swap-out construction, described in the next subsection.

\subsection{Swap-Based Realization of Conditional Skip}
To reduce the circuit depth and control overhead associated with conditionally applying a deep subroutine $U_B$, we implement an equivalent skip mechanism using a pair of controlled-SWAP operations. The idea is to redirect the action of $U_B$ away from the true data register $x_B$ whenever the skip condition is met. Specifically, we introduce a dummy register $d_B$ initialized to the state $\ket{0}^{\otimes n}$ and perform the following three steps:
\begin{enumerate}
    \item Apply a controlled-SWAP gate (Fredkin gate) between each corresponding pair of qubits in $x_B$ and $d_B$, with the ancilla $a$ as the control. This swaps the content of $x_B$ into $d_B$ when $a = 1$ and leaves $x_B$ unchanged when $a = 0$.
    \item Apply $U_B$ unconditionally to the register currently labeled $x_B$.
    \item Repeat the same controlled-SWAP operation to restore the original register assignment.
\end{enumerate}
Because $d_B$ is initialized to $\ket{0}^{\otimes n}$ and is assumed not to satisfy the condition encoded by $U_B$, the action of $U_B$ on $d_B$ is the identity. Thus, the net effect is that $U_B$ is applied if and only if $a = 0$. Formally, the full unitary transformation is given by
\begin{equation}
V_{\text{swap}} = \left( \ket{0}\bra{0}_a \otimes U_B + \ket{1}\bra{1}_a \otimes I \right),
\end{equation}
realized via conjugation of $U_B$ by controlled-SWAP layers:
\begin{equation}
V_{\text{swap}} = \left( \text{CSWAP}_a \right) \cdot \left( I \otimes U_B \right) \cdot \left( \text{CSWAP}_a \right).
\end{equation}
Each CSWAP decomposes into three elementary gates (two CX and one H), so the total overhead scales linearly with the width of $x_B$. Unlike direct multi-controlled constructions, this approach preserves depth efficiency while faithfully implementing the skip logic.

\subsection{Flag Setting and Ancilla Uncomputation}

After the conditional execution of $U_B$, the data register $x_B$ may contain an output state that satisfies a predetermined success condition. To coherently record whether this condition is met, we apply a unitary that flips the flag qubit $f_B$ if and only if $x_B$ is in a specific marked basis state $\ket{w}$. This is implemented using the projector $\Pi_B = \ket{w}\bra{w}$ acting on $\mathcal{H}_{x_B}$. The corresponding flag-setting unitary is
\begin{equation}
\begin{aligned}
U_{\text{flag},B}
&= I \otimes \Big[ (I - \Pi_B) \otimes I_{f_B} \\
&\qquad\quad + \Pi_B \otimes X_{f_B} \Big],
\end{aligned}
\end{equation}
where the identity extends over all registers not explicitly shown. This transformation acts trivially unless $x_B = w$, in which case it flips $f_B$ from $\ket{0}$ to $\ket{1}$. Equivalently, the flag $f_B$ is flipped if and only if the post-$U_B$ state of $x_B$ equals the marked bitstring $\ket{w}$.

Following flag-setting, we uncompute the ancilla qubit $a$, restoring it to the state $\ket{0}$. This is accomplished by reapplying the same Toffoli-style gate used to compute the skip condition $C \wedge f_A$:
\begin{equation}
\begin{aligned}
V_{\text{uncompute}}
&= V_{\text{AND}}^\dagger \\
&= \text{Toffoli}_{C,f_A \rightarrow a},
\end{aligned}
\end{equation}
with all other registers acted on by the identity. Because $a$ is not acted on by any operations between its initial computation and this uncomputation, and because $V_{\text{AND}}$ is unitary, this step exactly reverses the earlier entanglement and disentangles the ancilla from the rest of the system.

\subsection{Overall Unitary of One QSG Layer}
We now express the full unitary corresponding to a single application of the Quantum Skip Gate logic. Let $U_A$ and $U_B$ be two unitary subroutines acting on registers $(x_A, f_A)$ and $(x_B, f_B)$, respectively, and let $D$ denote an arbitrary post-processing unitary (such as Grover diffusion) acting on $(x_A, x_B)$. The full Hilbert space is decomposed as $\mathcal{H}_C \otimes \mathcal{H}_{x_A} \otimes \mathcal{H}_{x_B} \otimes \mathcal{H}_{f_A} \otimes \mathcal{H}_{f_B}$.
The overall QSG-layer unitary is given by
\begin{align}
U_{\text{QSG}} =\;&
\left( \Pi_{C=0} \otimes U_B U_A D \right) \nonumber \\
&+ \left( \Pi_{C=1} \otimes
     \left[ (I - \Pi_{f_A}) U_B \right. \right. \nonumber \\
&\left. \left. \quad +\; \Pi_{f_A} \cdot I \right] U_A D \right)
\end{align}
where $\Pi_{C=0} = \ket{0}\bra{0}_C$, $\Pi_{C=1} = \ket{1}\bra{1}_C$, and $\Pi_{f_A} = \ket{1}\bra{1}_{f_A}$ is the projector that signals success of subroutine $U_A$.
This unitary acts blockwise on the control qubit $C$. In the $C = 0$ branch, both $U_A$ and $U_B$ are applied unconditionally. In the $C = 1$ branch, the operator $U_B$ is coherently replaced by the identity $I$ whenever $f_A = 1$, implementing the skip. This formalizes the quantum-controlled omission of a subroutine based on internal logic, as performed by the Quantum Skip Gate.

\section{Embedding QSG in Grover Search}
\subsection{Operational Picture}
While the Quantum Skip Gate (QSG) is an application-independent primitive, its utility is illustrated well by the layered structure of Grover-style search algorithms. In this section, we present an explicit formulation of the QSG unitary when used in conjunction with Grover oracles and diffusion steps.
We label the full basis as $\ket{C, x_A, x_B, f_A, f_B}$ and let $U_A$ and $U_B$ denote two phase-oracles, with $D$ representing the usual Grover diffusion operator. An auxiliary register $d_B$ of $n$ qubits is introduced for the swap-out implementation; it is initialized and reset to $\ket{0}^{\otimes n}$ each iteration and therefore factors out of the effective unitary description. The overall unitary for one Grover iteration of the QSG-enhanced circuit is
\begin{align}
U_{\mathrm{QSG}} =\;&
\left( \Pi_{C=0} \otimes U_B U_A D \right) \nonumber \\
&+ \left( \Pi_{C=1} \otimes
     \left[ (I - \Pi_{f_A}) U_B \right. \right. \nonumber \\
&\left. \left. \quad +\; \Pi_{f_A} \cdot I \right] U_A D \right)
\end{align}
where $\Pi_{C=0} = \ket{0}\bra{0}_C$, $\Pi_{C=1} = \ket{1}\bra{1}_C$, and $\Pi_{f_A=i} = \ket{i}\bra{i}_{f_A}$. These projectors act on the control and flag subspaces and satisfy the standard properties
\begin{equation}
\begin{aligned}
\Pi_{f_A=0} + \Pi_{f_A=1} &= I, \\
\Pi_{f_A=i}\,\Pi_{f_A=j} &= \delta_{ij}\,\Pi_{f_A=i}.
\end{aligned}
\end{equation}
In the $C=0$ branch, both subroutines $U_A$ and $U_B$ are executed unconditionally. In the $C=1$ branch, $U_B$ is coherently replaced by the identity whenever $f_A = 1$, thereby realizing the skip. The overall unitary remains block-diagonal and norm-preserving, with coherent logic flow controlled entirely by quantum projectors.

To illustrate this more concretely, consider the case $n = 1$, where the skip logic affects only two qubits, $x_B$ and $f_A$. In this reduced subspace, the $C=1$ block of the overall unitary becomes

\begin{equation}
U_{C=1} =
\begin{pmatrix}
U_B U_A & 0 & 0 & 0 \\
0 & I U_A & 0 & 0 \\
0 & 0 & U_B U_A & 0 \\
0 & 0 & 0 & I U_A
\end{pmatrix}.
\end{equation}
This $4 \times 4$ matrix acts on the $(x_B, f_A)$ subspace. The remaining registers $C$, $x_A$, and $f_B$ are unaffected and factor out as identities, yielding a full operator of the form $I_{C,x_A,f_B} \otimes U_{C=1}$, which is $8 \times 8$ in total. The second and fourth rows of $U_{C=1}$, corresponding to $f_A = 1$, demonstrate explicitly that $U_B$ is coherently skipped when success is flagged by $U_A$.

\subsection{Grover Layering and Skip-Aware Iteration}
\begin{figure}[ht]
\centering
\includegraphics[width=\linewidth]{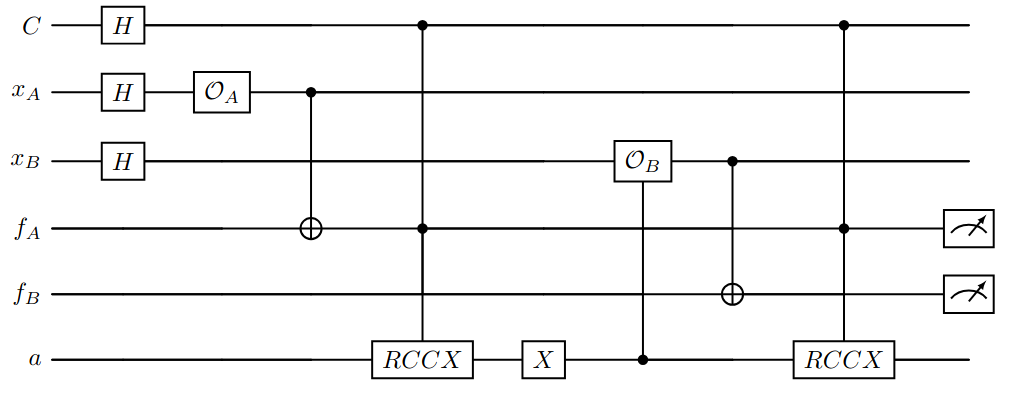}
\caption{High-level diagram of the Quantum Skip Gate applied within a Grover-style circuit. The control qubit $C$, data registers $x_A$ and $x_B$, flag qubits $f_A$, $f_B$, and ancilla $a$ are shown. Oracle $\mathcal{O}_A$ always runs, and $\mathcal{O}_B$ is conditionally executed depending on the conjunction $C \wedge f_A$, computed into $a$ via RCCX. Grover diffusion $D$ follows. This sketch omits low-level optimizations such as the swap-out construction.}
\label{fig:highlevel-grover}
\end{figure}

The QSG primitive may be embedded into Grover iterations of the form
\begin{equation}
U_{\text{layer}} = V \cdot U_B U_A D,
\end{equation}
where $V$ implements skip logic via conjugation or conditional branching. For instance, when swap-out logic is used, one may express
\begin{equation}
V = I - \ket{1}\bra{1}_C \otimes \Pi_A \otimes (I - S),
\end{equation}
where $\Pi_A = \ket{w}\bra{w}_{x_A}$ and $S$ is the identity on $x_B$, since skipping leaves the state unchanged. The unitary $V$ is block-diagonal and satisfies $[V, \Pi_A] = 0$, $[V, U_A] = 0$, so that amplitude amplification proceeds identically on the ``execute'' and ``skip'' branches except for the presence or absence of $U_B$. Iterating $k$ times yields the full search process:
\begin{equation}
U(k) = \left( U_{\text{layer}} \right)^k,
\end{equation}
where each layer includes skip-aware conditioning on the flag set by $U_A$. This embedding illustrates how coherent skip logic modifies the internal dynamics of Grover search without violating unitarity or coherence.
Figure~\ref{fig:highlevel-grover} provides a visual summary of the Quantum Skip Gate (QSG) control logic in the context of a Grover-style circuit. It highlights the core skip condition and the role of the ancilla qubit while deferring low-level details—such as the swap-out construction—to later sections.
A full Qiskit implementation of the QSG Grover iteration, including the swap-out logic and both QSG and fixed-order circuit constructors, is provided in Appendix~\ref{sec:appendix-code}.

\subsection{Low-Cost Implementation of the Skip Condition}
To implement the condition $a = C \wedge f_A$, a standard Toffoli gate would require six CX gates. We instead use Qiskit's \texttt{RCCXGate()}, a relative-phase Toffoli variant\cite{Maslov_2016}, which achieves the same classical logic using only three CX gates and a sequence of single-qubit gates. Although this gate introduces a known relative phase, the phase is unobservable in our setting, since the ancilla $a$ is later uncomputed. The specific construction used follows Qiskit’s implementation of \texttt{RCCXGate()}, which applies three CX gates along with $T$, $T^\dagger$, $H$, and $S$ rotations to reduce depth and preserve classical behavior.

\subsection{Swap-Out Realization of the Expensive Subroutine}
\label{subsec:swap_out_ub}
The original circuit places the verification oracle $U_B$ under an additional control so that it executes only when the ancilla $a = C \wedge f_A$ is $0$. Appending this extra control to every gate inside $U_B$ forces the transpiler to decompose the oracle into a large multi-controlled network, giving rise to the severe depth disparity with the fixed-order baseline.
Instead of controlling a heavyweight operator, we redirect its action. We use an $n$-qubit dummy register $d_B$ initialized to $\ket{0}^{\otimes n}$ and perform two rounds of controlled-SWAP (Fredkin) gates:
\begin{enumerate}
  \item If $a=1$, swap the data register $x_B$ with $d_B$; if $a=0$, leave $x_B$ in place.
  \item Apply $U_B$ \emph{unconditionally} to the lines currently labeled $x_B$.
  \item Repeat the same controlled-SWAP block to restore the original allocation.
\end{enumerate}
Because $U_B$ acts on $\ket{0}^{\otimes n}$ as the identity whenever that all-zero string is not a marked item, the composite unitary on $x_B$ and $a$ is identical to the intended ``run/skip'' behavior. Formally,
\begin{align}
&\left( \ketbra{0}{0}_a \otimes I + \ketbra{1}{1}_a \otimes S \right)
\left( I \otimes U_B \right) \nonumber \\
&\qquad\quad \cdot\,
\left( \ketbra{0}{0}_a \otimes I + \ketbra{1}{1}_a \otimes S \right) \nonumber \\
&= \ketbra{0}{0}_a \otimes U_B + \ketbra{1}{1}_a \otimes I,
\label{eq:swap_sandwich}
\end{align}
where $S$ swaps $x_B$ and $d_B$. Equation~\eqref{eq:swap_sandwich} replaces the multi-controlled version of $U_B$ with two $n$\,CSWAP layers plus the \emph{bare} oracle.
A textbook CSWAP decomposes into $2$\,CX $+ 1$\,H per target, so the pair of swap blocks costs $4n$\,CX in total. For $n=4$ and $R=30$ Grover repetitions this removes ${\sim}1.5\times10^{4}$ CX gates and ${\sim}5\times10^{4}$ single-qubit layers, bringing the QSG depth to within $10\%$ of the fixed-order circuit.
The ancilla $a$ remains untouched, so the subsequent $\texttt{RCCXGate()}$ uncomputes it exactly as in the original design. The dummy register is disentangled at the end of every Grover iteration, ensuring no coherence penalty accrues across iterations.

\section{Relation to Prior Work}
This work builds on quantum control of subroutines and resource-efficient oracle design in iterative algorithms.

\subsection{Controlled Subroutines and Skip Logic}
The ability to conditionally apply unknown operations is central to our circuit design. Zhou et al.~\cite{Zhou_2011} showed that it is possible to construct a controlled-$U$ gate even when $U$ is unknown, providing a critical building block for coherent skip logic. Friis et al.~\cite{Friis_2014} explored similar techniques in the context of quantum control of subroutine execution.
Wechs et al.~\cite{Wechs_2021} distinguished between classical dynamic circuits, where mid-circuit measurement determines future operations, and purely quantum-controlled logic. Our construction uses purely quantum control without measurement, avoiding circuit resets and preserving coherence across branches. It can be viewed as a hybrid between order-superposition ideas and early-exit logic from dynamic circuits.

\subsection{Closest Related Algorithms}
Liu et al.~\cite{liu2023experimentallydemonstratingindefinitecausal} demonstrated order-superposition techniques for single-shot decision problems. In contrast, our construction implements coherent conditional skipping within an iterative quantum algorithm, using quantum control to decide whether to invoke a costly subroutine. While we demonstrate this in Grover search, the skip logic itself is broadly applicable to any multi-round algorithm where conditional execution can yield resource savings.

\subsection{Query Complexity Limits}
Abbott et al.~\cite{Abbott_2024} rigorously showed that certain advanced control techniques do not improve asymptotic query complexity for total Boolean functions, so Grover’s $\Theta(\sqrt{N})$ scaling remains optimal. However, they noted that such techniques can yield constant-factor improvements in specific cases by lowering error probability for a fixed number of queries. Our results complement this view. While no scaling improvement is claimed, we demonstrate a substantial efficiency gain under realistic cost models.

\section{Experimental Results}
\subsection{Brisbane Hardware Sweep \emph{without} Swap-Out QSG}
We executed the Quantum Skip Gate (QSG) and a fixed-order Grover baseline on \texttt{ibm\_brisbane} for $n = 4$ data qubits, $k = 3$ Grover iterations, and two verification depths $R \in \{20, 30\}$. Each circuit was sampled with 4{,}000 shots. Table~\ref{tab:brisbane} summarizes the metrics extracted directly from the device run logs.
\begin{table}[ht]
\centering
\caption{
Performance of the Quantum Skip Gate (QSG) versus fixed-order Grover on \texttt{ibm\_brisbane}. Depth is the transpiled one-qubit depth; “ECR” counts native two-qubit gates. Efficiency $\eta$ is defined as $P_{\text{succ}}/\langle \#U_B \rangle$. Runs \#1 and \#2 use $R=20$ with different register orderings; run \#3 uses $R=30$.
}
\label{tab:brisbane}
\resizebox{\linewidth}{!}{%
\begin{tabular}{ccccccc}
\toprule
$R$ & Strategy & Depth & ECR & $P_{\text{succ}}$ & $\langle \#U_B \rangle$ & $\eta$ \\
\midrule
\multirow{2}{*}{20}
 & QSG & 52{,}514 & 14{,}789 & $0.7565 \pm 0.0068$ & 5.44 & 0.1391 \\
 & Fixed & 13{,}383 & 4{,}339 & $0.6347 \pm 0.0076$ & 6.00 & 0.1058 \\
\cmidrule{2-7}
 & Fixed & 13{,}571 & 4{,}501 & $0.6375 \pm 0.0076$ & 6.00 & 0.1062 \\
 & QSG & 51{,}966 & 14{,}846 & $0.7662 \pm 0.0067$ & 4.49 & 0.1706 \\
\midrule
30 & QSG & 74{,}534 & 20{,}924 & $0.1857 \pm 0.0061$ & 5.65 & 0.0329 \\
   & Fixed & 16{,}349 & 5{,}395 & $0.6048 \pm 0.0077$ & 6.00 & 0.1008 \\
\bottomrule
\end{tabular}
}
\end{table}
At $R = 20$, corresponding to runs \#1 and \#2, the QSG circuit achieved $31\%$ and $61\%$ higher efficiency than the fixed-order baseline. This improvement stemmed from a $10$--$20\%$ increase in success probability while simultaneously reducing the number of calls to the expensive oracle $U_B$ by $9$--$25\%$. In contrast, at $R = 30$ (run \#3), the QSG circuit reached a depth of approximately $7.5 \times 10^4$ one-qubit layers, and the accumulation of two-qubit noise overwhelmed the gains from skipping, causing the efficiency to drop to roughly one-third that of the fixed-order baseline.
Interpolating between these data points suggests that the crossover point—where QSG transitions from beneficial to detrimental on this hardware—lies in the range $25 \lesssim R \lesssim 30$.
These results prompted a deeper investigation into the source of QSG's performance degradation at higher oracle depths. The dominant factor was the circuit depth overhead incurred by placing the entire oracle $U_B$ under control, which forced the transpiler to decompose $U_B$ into a large multi-controlled gate network. To mitigate this cost, we developed a depth-optimized “swap-out” implementation, which avoids controlling $U_B$ directly by conditionally redirecting the data register through a dummy wire. This construction preserves the skip logic while significantly reducing depth, enabling QSG to remain competitive even as oracle complexity grows.

\subsection{Noisy-Hardware Benchmarks \emph{with} Swap-Out QSG on \texttt{ibm\_sherbrooke}}
We emulate the 127-qubit \emph{ibm\_sherbrooke} processor with Qiskit~Aer’s noise model and compare the Quantum Skip Gate (QSG) circuit against the fixed-order Grover baseline. Simulation parameters are identical across runs, with $n=4$ data qubits per oracle, $k=3$ Grover iterations, and $1000$ shots per circuit.
\begin{table}[ht]
\centering
\caption{Performance at three oracle costs $R$ (depth of the expensive verification oracle $U_B$). Depth and two-qubit ECR counts are those of the transpiled circuits; efficiencies are defined as $\,P_{\text{succ}}/\,\langle \#U_B \rangle\,$.}
\label{tab:sherbrooke-bench}
\resizebox{\linewidth}{!}{%
\begin{tabular}{@{}l c r r r r c@{}}
\toprule
$R$ & Circuit & Depth & ECR & $P_{\text{succ}}$ & $\langle \#U_B \rangle$ & Efficiency \\
\midrule
\multirow{2}{*}{25} & QSG & 12\,198 & 4\,232 & 0.751 & 4.53 & \textbf{0.166} \\
                    & Fixed & 15\,221 & 4\,912 & 0.739 & 6.00 & 0.123 \\[2pt]
\multirow{2}{*}{30} & QSG & 14\,819 & 5\,357 & 0.737 & 4.53 & \textbf{0.163} \\
                    & Fixed & 16\,017 & 5\,299 & 0.697 & 6.00 & 0.116 \\[2pt]
\multirow{2}{*}{35} & QSG & 15\,867 & 5\,141 & 0.721 & 4.57 & \textbf{0.158} \\
                    & Fixed & 18\,473 & 6\,379 & 0.645 & 6.00 & 0.108 \\
\bottomrule
\end{tabular}%
}
\end{table}
For all oracle depths tested, QSG maintains a smaller physical depth and consistently skips $\approx 25\%$ of $U_B$ calls ($4.5$ versus $6$ on average). Despite similar entangling-gate counts, QSG achieves higher success probabilities and outperforms the fixed-order circuit in efficiency by $35\%$, $40\%$, and $45\%$ for $R=25,30,35$ respectively. These results confirm that coherent skip logic offers a tangible advantage on realistic noisy hardware at high oracle cost.

\subsection{Hardware Run on \texttt{ibm\_brisbane} with Swap-Out QSG at R = 10}
With the swap-out construction in place, QSG is now $\approx24\%$ shallower than the fixed sequence even at the lower oracle depth $R=10$. The circuit skips $25\%$ of $U_B$ invocations on average and more than doubles the success probability, yielding an efficiency nearly \emph{three times} higher than the baseline. Together with the Sherbrooke-noise simulations at $R\ge25$, this hardware run confirms that the Quantum Skip Gate delivers a performance gain across the full range of oracle costs tested, from shallow to deeply nested verification stages.
\begin{table}[ht]
\centering
\caption{Real-device results for a shallower verification oracle ($R=10$) on \texttt{ibm\_brisbane}. Depth and ECR counts are for the transpiled circuits compiled to the machine’s native gate set. Each circuit was executed with $1000$ shots. Efficiency is $P_{\text{succ}}/\langle \#U_B \rangle$.}
\label{tab:brisbane-R10}
\resizebox{\linewidth}{!}{%
\begin{tabular}{@{}l r r r r c@{}}
\toprule
Circuit & Depth & ECR & $P_{\text{succ}}$ & $\langle \#U_B \rangle$ & Efficiency \\
\midrule
QSG & 8\,058 & 2\,840 & 0.763 & 4.49 & \textbf{0.170} \\
Fixed & 10\,583 & 3\,595 & 0.355 & 6.00 & 0.059 \\
\bottomrule
\end{tabular}%
}
\end{table}

\section{Conclusion}
The Quantum Skip Gate enables practical coherent conditional execution of expensive subroutines in gate-model quantum algorithms. By allowing a costly oracle $U_B$ to be skipped coherently when a cheaper oracle $U_A$ has already succeeded, QSG provides effective resource management in iterative algorithms.

A practical challenge was the depth overhead of controlling a deep oracle. We solved this with a swap-out implementation that reroutes the data through a dummy register rather than adding an extra control to every gate inside $U_B$. This change keeps the physical depth nearly constant as the oracle cost $R$ grows, and restores the theoretical advantage of skip logic across the full range we tested.

On real \texttt{ibm\_brisbane} hardware at $R=10$ the swap-optimized QSG is $24\%$ shallower than the fixed-order circuit, more than doubles the raw success probability, and delivers almost three times the success-per-oracle efficiency. Noise-model simulations based on the $127$-qubit \texttt{ibm\_sherbrooke} processor show that the advantage persists and even grows at higher oracle depths ($R=25,30,35$), reaching efficiency gains of $35$--$45\%$ while maintaining higher success probabilities in every case.

These results demonstrate that coherent quantum control can cut runtime cost and mitigate noise in near-term devices. Future work will scale the QSG to larger databases and deeper verification oracles and explore its use in other applications, such as adaptive phase-estimation protocols. Additional applications may include quantum metrology, resource-aware variational circuits, or fault-tolerant schemes where selective skipping reduces exposure to decoherence.

\bibliographystyle{quantum}
\bibliography{references}

\appendix
\section{Qiskit Code Example Implementation (Illustrative)}
\label{sec:appendix-code}
This appendix provides illustrative Qiskit implementations of the two core circuit builder functions used in this work. Figure~\ref{lst:build-qsg} defines the Quantum Skip Gate (QSG) Grover-layer constructor, which implements conditional oracle skipping using the swap-out technique. Figure~\ref{lst:build-fixed} shows the corresponding fixed-order Grover-layer builder for comparison, in which both oracles are always executed.
\begin{figure*}[t!]
\begin{lstlisting}[style=qsg,
    caption={Builder function for the Quantum Skip Gate (QSG) circuit using swap-out logic.},
    label={lst:build-qsg}]
def build_qsg(reps):
    """Quantum Skip Gate (QSG) Grover layer using swap out trick."""
    NQ = 3*n + 4
    qc = QuantumCircuit(NQ, 2)
    C = 0
    xA = list(range(1, n+1))
    xB = list(range(n+1, 2*n+1))
    fA, fB = 2*n+1, 2*n+2
    anc = 2*n+3
    dB = QuantumRegister(n, "dB"); qc.add_register(dB)
    qc.h(xA + xB + [C])
    for _ in range(k):
        qc.append(phase_oracle(n, OA_mask, "O_A"), xA)
        set_flag(qc, xA, fA, OA_mask)
        qc.append(RCCXGate(), [C, fA, anc])
        for i in range(n):
            qc.cswap(anc, xB[i], dB[i])
        qc.append(expensive_oracle(n, OB_mask, reps), xB)
        for i in range(n):
            qc.cswap(anc, xB[i], dB[i])
        set_flag(qc, xB, fB, OB_mask)
        qc.append(RCCXGate(), [C, fA, anc])
        diffusion(qc, xA + xB)
    qc.measure(fA, 0)
    qc.measure(fB, 1)
    return qc
\end{lstlisting}
\end{figure*}
\begin{figure*}[t!]
\begin{lstlisting}[style=qsg,
    caption={Builder function for the fixed-order Grover circuit.},
    label={lst:build-fixed}]
def build_fixed(reps):
    """Fixed order Grover (always run both oracles)."""
    NQ = 2*n + 3
    qc = QuantumCircuit(NQ, 2)
    xA = list(range(n))
    xB = list(range(n, 2*n))
    fA, fB = 2*n, 2*n+1
    qc.h(xA + xB)
    for _ in range(k):
        qc.append(phase_oracle(n, OA_mask, "O_A"), xA)
        set_flag(qc, xA, fA, OA_mask)
        qc.append(expensive_oracle(n, OB_mask, reps), xB)
        set_flag(qc, xB, fB, OB_mask)
        diffusion(qc, xA + xB)
    qc.measure(fA, 0)
    qc.measure(fB, 1)
    return qc
\end{lstlisting}
\end{figure*}
\end{document}